# eFIN: Enhanced Fourier Imager Network for generalizable autofocusing and pixel super-resolution in holographic imaging

Hanlong Chen, Luzhe Huang, Tairan Liu, Aydogan Ozcan*

*Abstract*—The application of deep learning techniques has greatly enhanced holographic imaging capabilities, leading to improved phase recovery and image reconstruction. Here, we introduce a deep neural network termed enhanced Fourier Imager Network (eFIN) as a highly generalizable framework for hologram reconstruction with pixel super-resolution and image autofocusing. Through holographic microscopy experiments involving lung, prostate and salivary gland tissue sections and Papanicolau (Pap) smears, we demonstrate that eFIN has a superior image reconstruction quality and exhibits external generalization to new types of samples never seen during the training phase. This network achieves a wide autofocusing axial range of Δz ~ 350 μm, with the capability to accurately predict the hologram axial distances by physics-informed learning. eFIN enables 3× pixel super-resolution imaging and increases the space-bandwidth product of the reconstructed images by 9-fold with almost no performance loss, which allows for significant time savings in holographic imaging and data processing steps. Our results showcase the advancements of eFIN in pushing the boundaries of holographic imaging for various applications in e.g., quantitative phase imaging and label-free microscopy.

## I. Introduction

Digital holography has gained significant attention in recent years as a powerful tool for microscopic imaging, capable of reconstructing the complex optical fields of samples without the need for staining, fixation, or labeling of specimen [1]–[11]. A key challenge in digital holography is the recovery of the missing phase information from the recorded intensity-only hologram, which can be addressed through computational approaches, such as iterative algorithms based on physical wave propagation [12]–[21] and, more recently, deep neural networks. The recent deep learning-based approaches utilize neural networks to reconstruct the complex sample field from one or more holograms in a single forward inference step, providing advantages over traditional methods, including faster reconstruction speed, improved signal-to-noise ratio, and extended depth-of-field [22]–[44]. In addition, these methods have successfully achieved *internal generalization*, defined as the ability of the deep neural network to reconstruct holograms of new, unseen objects belonging to the *same* sample type used during the training. Recently, Chen *et al*. introduced the Fourier Imager Network (FIN),[45] a deep neural network that utilizes a Spatial Fourier Transform module to learn the mapping between the input holograms and the complex sample fields through a global receptive field. The FIN succeeded in performing *external* generalization by reconstructing holograms of new, unseen samples belonging to *different* sample types that were never seen during the training. However, FIN lacks the ability to perform autofocusing, meaning that a single trained FIN network can only reconstruct a sample's complex field using holograms captured at specific, known axial distances.

In this paper, we present the enhanced Fourier Imager Network (eFIN), a deep neural network for end-to-end phase retrieval and holographic image reconstruction that achieves both pixel super-resolution and auto-focusing – performed at the same time through its rapid inference. While derived from FIN, eFIN differs significantly in its architecture by utilizing a shallow U-Net in its Dynamic Spatial Fourier Transform (SPAF) module, and provides additional degrees of freedom to integrate the capabilities of pixel super-resolution and auto-focusing in a single neural network. Specifically, eFIN stands out in its ability to perform autofocusing over a wide axial range (Δz) of 350 μm and accurately predicts the axial distances of the input holograms through physics-informed learning, obviating the need for ground truth axial distance values. Moreover, eFIN exhibits superior image reconstruction quality on both internal and external generalization tasks and is able to achieve 3× pixel super-resolution (PSR) with minimal performance compromise, saving a significant amount of time in the imaging and data processing steps. Stated differently, eFIN is trained (1) to perform pixel super-resolution hologram reconstruction revealing both the amplitude and phase images of the input specimen with higher resolution; (2) to perform axial autofocusing since the sample-to-sensor distances are unknown and vary from input to input; and (3) reveal the axial distances of the input holograms in addition to phase retrieval, hologram PSR reconstruction and autofocusing.

To experimentally validate the success of eFIN, we trained it

The Ozcan Research Group at UCLA acknowledges the support of NSF PATHS-UP.

* Correspondance to: ozcan@ucla.edu

H. Chen, L. Huang, T. Liu and A. Ozcan are with the Electrical and Computer Engineering Department, Bioengineering Department, California Nano Systems Institute (CNSI), University of California, Los Angeles, CA 90095 USA, and also with the California NanoSystems Institute, University of California, Los Angeles, CA 90095 USA (email: hanlong@g.ucla.edu; lzhuang0324@g.ucla.edu; liutr@g.ucla.edu; ozcan@ucla.edu)

using the holograms of human lung tissue samples (histopathology slides) and blindly tested its inference, image reconstruction and autofocusing performance on Pap smears, human lung, prostate and salivary gland tissue sections (all of which were never seen during the training phase). The superior PSR image reconstruction and autofocusing performance of eFIN represents a major advancement in deep learning-based holographic imaging, presenting opportunities to greatly improve the efficiency and speed of high-resolution holographic microscopy. The versatility of eFIN also allows for its applications in various incoherent imaging modalities, making it a promising tool for a wide range of imaging and microscopy applications.

## II. RESULTS AND DISCUSSIONS

### A. The architecture of eFIN

eFIN is a deep neural network that performs end-to-end phase recovery and holographic image reconstruction with the ability to autofocus over a wide axial range and achieve pixel super-resolution image reconstruction, as shown in Fig 1. Additionally, a sub-network within eFIN, named Z-Net, is trained to blindly predict the sample-to-sensor axial distances of the input holograms without any a priori information.

The input of the network is a sequence of low-resolution intensity-only in-line holograms [46], captured at $M$ different sample-to-sensor distances, i.e., $z_1 ... , z_M$ (see the Methods section). The outputs are the real and imaginary parts of the pixel super-resolved reconstructed complex field of the sample, as well as the sample-to-sensor distances of the corresponding input holograms. The high-resolution ground truth (GT) complex-valued images for the supervised learning portion of our training are obtained using an iterative multi-height phase retrieval (MH-PR) algorithm [17], [18] using pixel super-resolved holograms acquired at eight different axial distances ($M = 8$) (see the Methods section).

### B. Pixel super-resolution hologram reconstruction and autofocusing performance of eFIN

To showcase the superior performance of eFIN, we trained the network using human lung tissue samples (see the Methods). During each training iteration, the network was fed two low-resolution holograms (i.e., $M = 2, k = 3$) captured at *randomly chosen* and *unknown* sample-to-sensor axial distances, $z$, within a range that covers 300 μm to 650 μm, in increasing order, where $k = 3$ stands for the targeted pixel super-resolution factor as illustrated in Fig. 2(a). Stated differently, each low-resolution in-line hologram has $k^2 - fold$ smaller number of pixels, yielding a resolution loss due to undersampling; for a unit magnification holographic on-chip microscope [47] this is equivalent to using an optoelectronic image-sensor with $k - fold$ larger pixel width/period, resulting in $k^2 - fold$ fewer pixel count (for example, $k^2 = 9$ in Fig. 2).

To first demonstrate its internal generalization, the eFIN network, after its training, was tested using new low-resolution input holograms of lung tissue samples from new patients (never seen before) acquired at various combinations of unknown $z$ distances; the resulting pixel super-resolution hologram reconstructions of eFIN are shown in Fig. 2(b). For comparison, we also present the results of the MH-PR algorithm in Fig. 2(c), which used the same low-resolution holograms as its input, but with the actual axial distances of each hologram (the $z$ values) known. A direct comparison of these results shows that our eFIN model delivers better hologram reconstruction quality across a wide range of sample-to-sensor axial distances without knowing the hologram $z$ values, i.e., performing both PSR image reconstruction and autofocusing at the same time. Additionally, eFIN was able to output the axial positions of the input holograms with a high degree of precision (see Fig. 2b). The yellow zoom-in areas in Fig. 2 further highlight the superior hologram reconstruction quality of eFIN.

To better understand the advantages of eFIN, we employed the same model weights as in Fig. 2 to assess its performance in detail. Figure 3(a) summarizes the eFIN image reconstruction performance values for the amplitude structural similarity index (SSIM), phase SSIM, and axial distance (z) prediction mean absolute error (MAE) for both internal generalization (*tested on the same type of samples as in the training phase but from new patients*) and external generalization (*tested on new types of samples never seen before*). Without the knowledge of the input low-resolution hologram axial distances, i.e., $z_1$ and $z_2$, eFIN was able to robustly reconstruct the sample complex field using $M = 2$ low-resolution in-line holograms ($k = 3$) randomly captured at sample-to-sensor z distances ranging from 300 μm to 650 μm, while also accurately predicting the sample-to-sensor distances of the input holograms for different combinations of $z$ values. Moreover, eFIN exhibits outstanding *external* generalization capability in its PSR image reconstruction and autofocusing performance, successfully reconstructing different types of samples (Pap smears, human prostate and salivary gland tissue sections) it has never seen before (see Figs. 3-4).

To further contrast the performance of eFIN with traditional hologram reconstruction methods, we also quantified the reconstruction quality of MH-PR algorithm. In this case, to provide a fair comparison against eFIN, we did not provide the accurate axial distances (z values) of the input holograms to MH-PR. Instead, two fixed holograms captured at $z_1 = 380 \mu m$ and $z_2 = 560$ μm, along with different combinations of varying axial distances ($z_1$ and $z_2$) were fed into the MH-PR algorithm to simulate the scenario where the precise axial distances of the input/acquired holograms are unavailable (which is frequently the case, especially in relatively inexpensive holographic imaging hardware). As shown in Fig. 3(b), the classical MH-PR algorithm cannot accurately reconstruct the sample field using inaccurate hologram axial distances; for comparison, the eFIN autofocusing results are shown in Fig. 3(a).

### C. eFIN reconstruction performance as a function of the PSR factor ($k$)

To further investigate the PSR hologram reconstruction performance of eFIN, we individually trained six different eFIN networks with the PSR factor $k$ ranging from 1 to 6. All the eFIN models were trained using only human lung tissue samples, with two input holograms ($M=2$) acquired at randomly selected and unknown sample-to-sensor distances from 300 μm to 650 μm. For the PSR factor $k=3$, we used the same model weights as in Figs. 2 and 3. These six different eFIN networks were tested with different combinations of $z_1$ and $z_2$, ranging from 300 μm, 400 μm, 500 μm and 600 μm, and they were blindly evaluated on human lung tissue sections (internal generalization) and Pap smear samples, human prostate and salivary gland thin tissue sections (external generalization). The mean SSIM values as a function of the aimed PSR factors ($k$) are plotted in Fig 4(a), and the reconstructed sample fields and their corresponding image metrics are reported in Fig 4(b). These results confirm the outstanding PSR performance of eFIN, revealing that the reconstruction quality is not compromised much, even when using low-resolution input holograms with $k^2=9$ or $k^2=16$ times fewer pixels compared to the $k=1$ case (see Fig. 4).

*D. eFIN model size and inference time*

We compared the model size and the inference (image reconstruction) time of eFIN, MH-PR, and the standard FIN in Table 1. By utilizing a shallow U-Net in the Dynamic SPAF module, eFIN is able to generate optimized weights of a linear transformation dynamically based on the input features, rather than relying on fixed, input-independent weights (see the Method section for details). This significantly reduces the model size of eFIN and enables its autofocusing and PSR capabilities. However, the U-Net components used within eFIN require more calculation time, resulting in a slightly longer PSR image inference time (~0.85 s/mm$^2$) for eFIN compared to FIN, even though eFIN has a model size that is approximately half of FIN.

TABLE I

|  | Number of trainable parameters | Inference time ($s/mm^2$) |
|---|---|---|
| eFIN (M=2) | 5.7M | 0.85 |
| FIN (M=2) | 11.5M | 0.51 |
| MH-PR (M=2) | N/A | 10.36 |

Model size and image reconstruction (inference) time comparison between eFIN, FIN and MH-PR, reported in seconds per 1 mm$^2$ sample field of view. Note that both MH-PR and FIN require the axial distances of the input holograms to be known a priori.

It is noteworthy that eFIN can perform pixel super-resolution and autofocusing with minimal image quality compromise using low-resolution input holograms, as shown in Fig. 4(a) for PSR factors $k=1,2,$ and 3. This allows eFIN to process raw holograms captured directly from a CMOS/CCD imager and output high-resolution sample fields (phase and amplitude) at the sub-pixel level. In contrast, MH-PR and FIN require pixel super-resolved holograms with known axial distances as input, which require additional hologram acquisition and data processing to perform PSR. These additional hologram/image acquisition and data processing steps needed for MH-PR and FIN are significantly more time-consuming compared to the inference times reported in Table 1.

III. METHODS

*A. Sample preparation and holographic imaging*

In this study, the human samples were obtained from deidentified and existing specimens collected before this study. UCLA Translational Pathology Core Laboratory (TPCL) collected and prepared prostate, salivary gland, and lung tissue slides, and the UCLA Department of Pathology supplied the Pap smear slides. To capture raw holograms of these specimens, a custom-designed lens-free in-line holographic microscope [17], [46] was utilized, where the light source aperture, the sample, and the CMOS image sensor were arranged vertically and positioned in the corresponding order. The illumination source was a broadband laser (WhiteLase Micro, NKT Photonics) equipped with an acousto-optic tunable filter, emitting 530 nm light, which was placed at a distance of approximately 10 cm above the sample. The CMOS image sensor (IMX081, Sony) was mounted on a 6-axis stage (MAX606, Thorlabs) and positioned at a distance ranging from 300 μm to 650 μm from the sample. Its lateral and axial shifts were controlled by a computer, running a customized LabVIEW program for automatic hologram capture.

*B. Dataset pre-processing*

To obtain the high-resolution target (GT) complex fields, we employed an iterative MH-PR algorithm, which used M=8 pixel super-resolved holograms (see Fig 2(a)), in addition to an autofocusing algorithm [48] providing the estimated sample-to-sensor distances for each acquired hologram. This iterative algorithm converged within 100 iterations. The high-resolution holograms were generated using a pixel super-resolution algorithm [49], which reduced the effective pixel size of the resulting super-resolved holograms to 0.37 μm. The raw in-line holograms were automatically captured at 6 × 6 lateral positions with sub-pixel shifts using the programmed 6-axis mechanical scanning stage.

The resulting high-resolution holograms and the target complex fields (GT) were cropped into 512 × 512 patches and divided into non-overlapping training, validation and testing sets with approximately 600, 100, and 100 FOVs for each sample type, respectively. Each image dataset only includes FOVs from different patients to ensure diversity. The low-resolution holograms were generated through pixel-binning using different PSR factors ranging from k=2 to k=6. The input low-resolution holograms to eFIN were ordered with increasing axial distances.

*C. Network structure*

The eFIN network structure is depicted in Fig 1(b). The dense links, inspired by the Dense Block in DenseNet [50], enable an efficient flow of information from the input layer to the output layer. Each output tensor of the Dynamic SPAF Group is appended and fed to the subsequent Dynamic SPAF Groups, resulting in the gradual accumulation of feature maps rather than a constant channel size throughout the network, thereby significantly reducing the number of parameters in the initial Dynamic SPAF Groups.

The Dynamic SPAF Modules of eFIN exploits a shallow U-Net [51] to dynamically generate weights $W$ for each input tensor, rather than using trainable weights fixed during the testing phase. The input tensor is first transformed into the spatial frequency domain using the two-dimensional fast Fourier transform (FFT), and a half window size $h$ is applied to filter out the high-frequency signals. Then, a linear transformation is performed as

$$F'_{j,u,v} = \sum_{i=1}^{c} F_{i,u,v} \cdot W_{j,u,v}$$
$$u, v = 0, \pm 1, \ldots, \pm h, \quad j = 1, \ldots, c \quad (1).$$

where $c$ is the channel number, $W \in \mathbb{R}^{c, 2h+1, 2h+1}$ represents the dynamically generated weights by U-Net, $F \in \mathbb{C}^{c, 2h+1, 2h+1}$ is the frequency components after the low-pass filter, and $F'$ is the transformed signal. $W$ is generated by feeding the absolute value of $F$ into a shallow U-Net. With this mechanism, the eFIN network can individually and uniquely adapt its weights to the input tensor, enabling autofocusing and pixel super-resolution capabilities.

The Z-Net (see Fig. 1) is a convolutional neural network, which takes in the outputs of the shallow U-Net from all the Dynamic SPAF Modules and predicts the (unknown) axial positions of the input holograms that were fed into eFIN. Rather than providing the ground truth values of $z$ for training the Z-Net by supervised learning, we instead designed a physics-informed learning framework, which aims to minimize the differences between the input low-resolution holograms and the intensity of the propagated fields using the ground truth complex field with the Z-Net predicted axial positions $\hat{z}$ (see Fig 1(a)). Compared to supervised learning, our physics-informed approach eliminates the need for the ground truth values of $z$, the axial distances of the input holograms, and also avoids introducing additional errors in training.

*D. Network implementation*

The network was implemented using PyTorch package [52] in Python. The training loss, as shown in Fig. 1(a), consists of two components: $Loss_s$ and $Loss_h$. $Loss_s$ represents the supervised loss for the reconstructed output complex field of the specimen, while $Loss_h$ represents the physics-informed loss for the Z-Net only.

The supervised loss $Loss_s$ is a weighted sum of MAE loss and the Fourier domain MAE (FDMAE) loss [45], [53]:

$$Loss_s = \alpha L_{MAE} + \beta L_{FDMAE} \quad (2).$$

where $\alpha$ and $\beta$ were empirically set to 2 and 0.007, respectively. These two terms can be expressed as:

$$L_{MAE} = \frac{\sum_{i=1}^{n}|y_i - \hat{y}_i|}{n} \quad (3).$$

$$L_{FDMAE} = \frac{\sum_{i=1}^{n}|\mathcal{F}(y)_i - \mathcal{F}(\hat{y})_i|}{n} \quad (4).$$

where $y$ is the GT, $\hat{y}$ is the output of the network, n is the total number of pixels, and $\mathcal{F}$ represents the two-dimensional FFT operator.

The physics-informed loss $Loss_h$ in Z-Net is the sum of the MAE values between the input low-resolution holograms and the holograms generated by propagating the GT complex fields to the planes located at the Z-Net predicted distances $\hat{z}_i, i = 1, \ldots, M$, i.e.,

$$Loss_h = \sum_{i=1}^{M} \left( \frac{\sum_{j=1}^{n}|x_{i,j} - PB(|FSP(y, \hat{z}_i)|, k)_j|}{n} \right) \quad (5).$$

where $x$ is the input low-resolution holograms, $y$ is the GT complex field of a sample, n is the total number of pixels, FSP is the free space propagation operation, $PB(\cdot, k)$ represents the $k - fold$ pixel-binning operation, and $M$ is the number of input holograms.

In the training phase, we used a computer with an Intel W-2195 CPU and Nvidia RTX 2080 Ti GPUs. The network was optimized iteratively using the AdamW optimizer [54] with a cosine annealed warm restart learning rate scheduler [55]. The best model was selected using the validation set, which only contains samples of the same type as those used in the training set. In the testing phase, the average inference time of the eFIN network is $0.85 \ s/mm^2$ on the same computer with a single RTX 2080 Ti GPU utilized.

ACKNOWLEDGMENTS

The Ozcan Research Lab at UCLA acknowledges the support of NSF PATHS-UP.

(a)

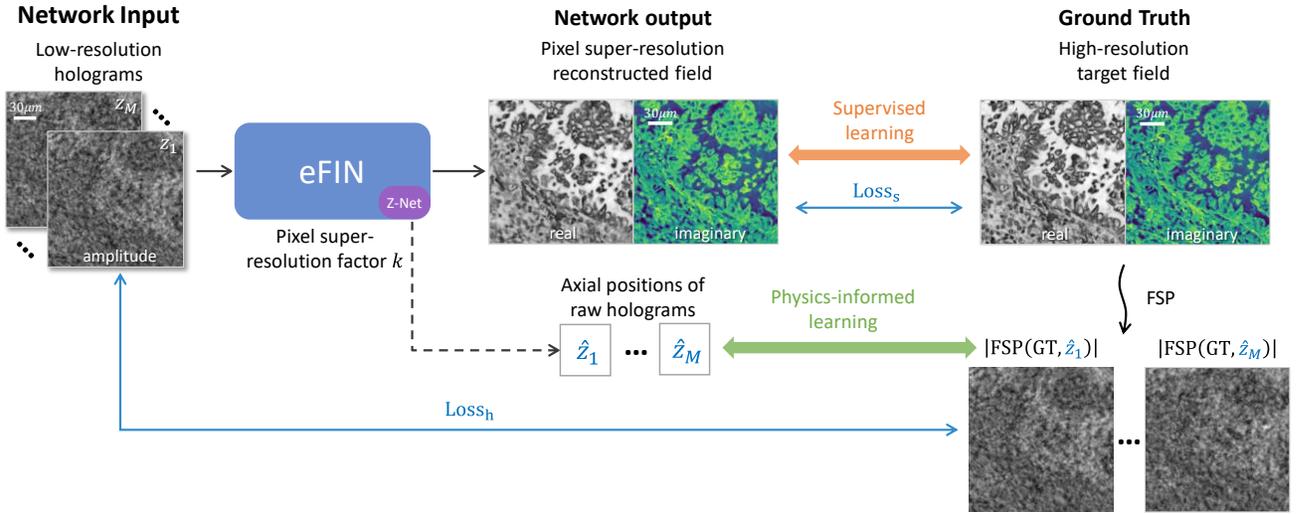

(b)

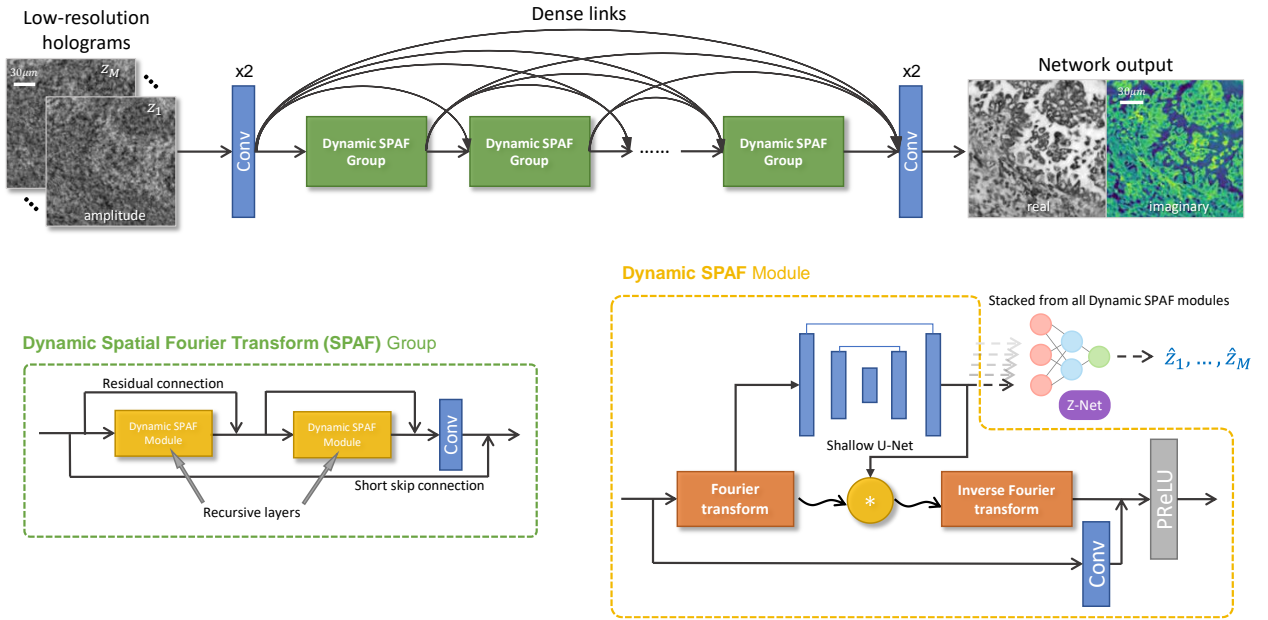

**Figure 1. Hologram reconstruction and autofocusing framework of eFIN.** (a) The input low-resolution holograms are captured at different sample-to-sensor distances $z_i$, $i = 1, ..., M$. $Loss_s$ is the supervised loss function derived from the ground truth target field, while $Loss_h$ is generated from physics-informed learning and applies only to Z-Net. (b) The architecture of the eFIN network, with Z-Net being a sub-network of eFIN. The outputs from the U-Net of all Dynamic SPAF modules are collected and fed into Z-Net for predicting the axial distances ($\hat{z}$) of the input holograms.

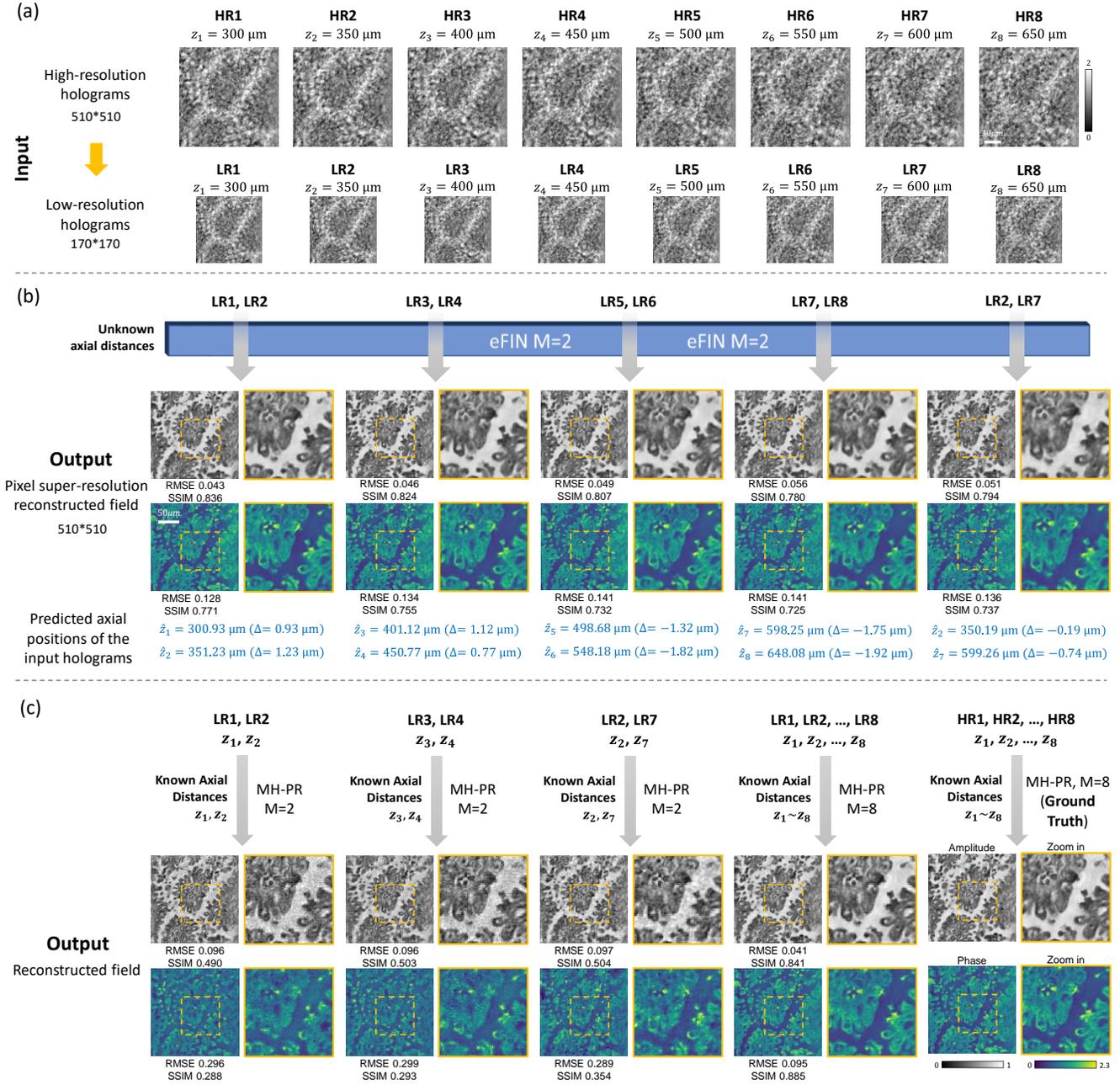

**Figure 2. Pixel super-resolved hologram reconstruction and autofocusing performance of eFIN.** (a) The low-resolution input holograms were obtained using 3× downsampling (by pixel-binning). (b) The output complex-valued fields were generated using the trained eFIN network ($M = 2$) with a pixel super-resolution factor of $k = 3$. (c) The output complex-valued fields of MH-PR algorithm, using the same low-resolution input holograms ($M = 2$), result in lower-quality reconstructions even though it uses the known axial distances for the input holograms.

(a)

**eFIN (M=2) performance matrices**

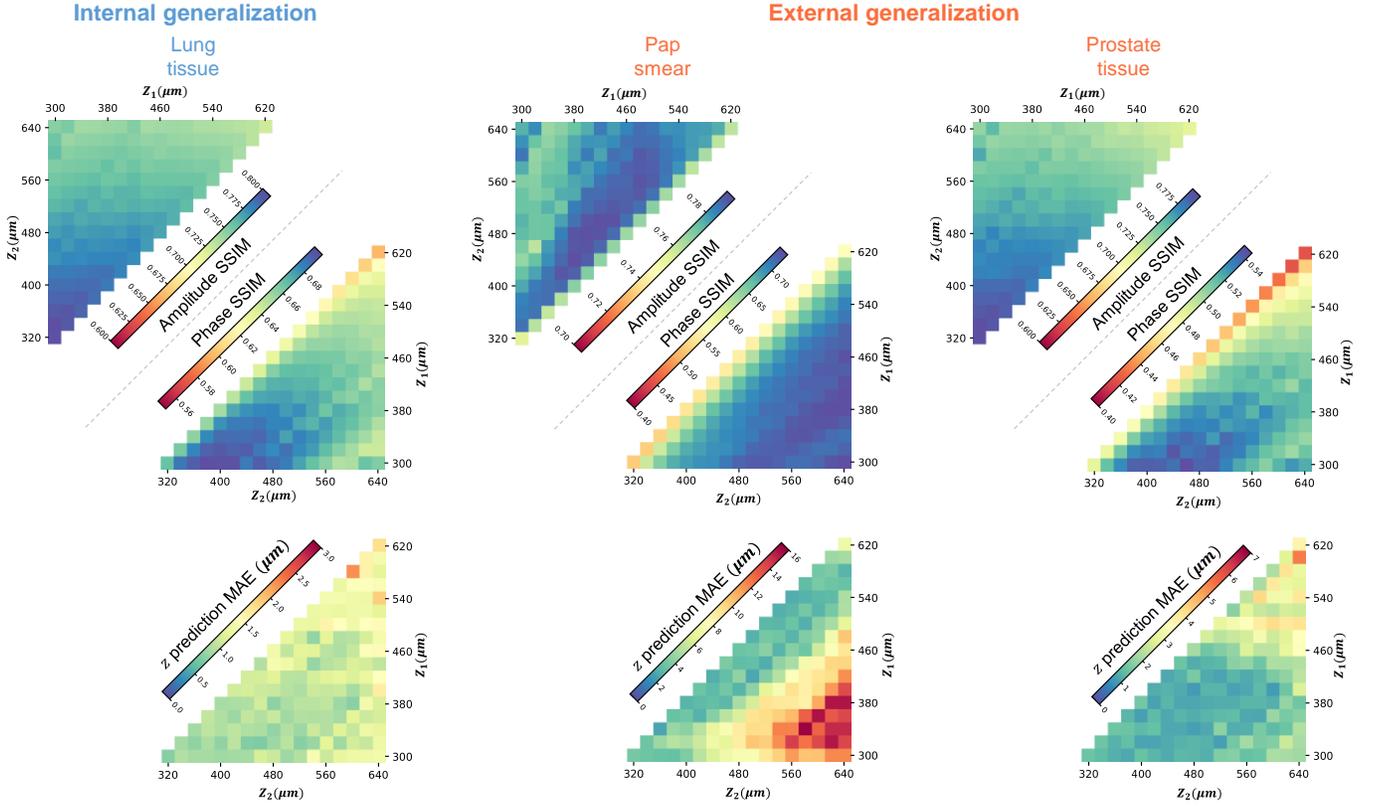

(b)

**MH-PR (M=2) performance matrices**

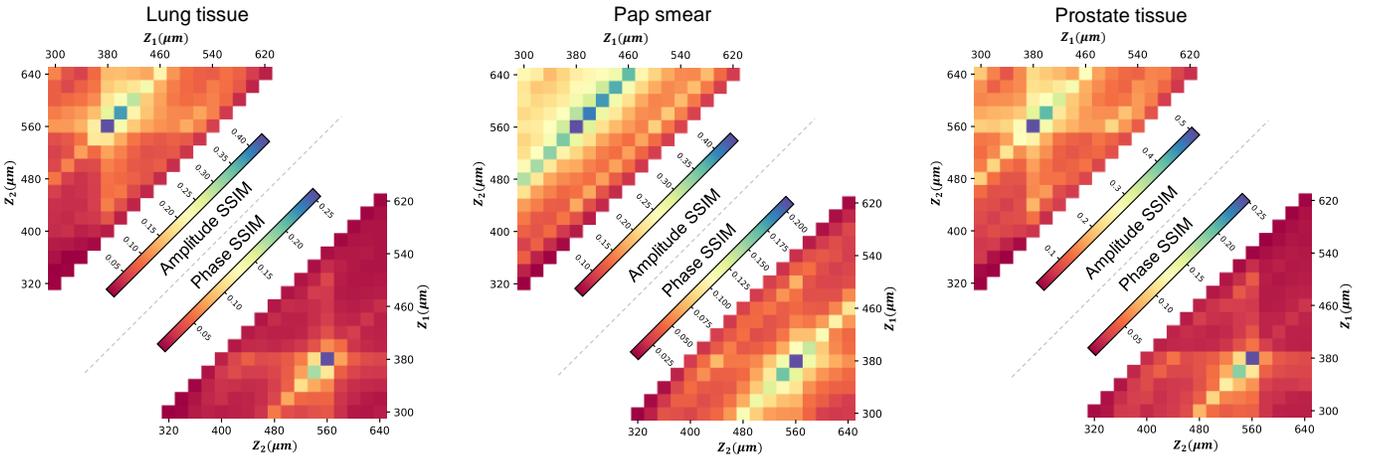

**Figure 3. Image reconstruction performance of eFIN and MH-PR.** (a) The amplitude and phase SSIM values of the reconstructed complex-valued fields and the MAE values of the predicted axial distances using the eFIN network ($M = 2, k = 3$). (b) The image reconstruction performance matrices of MH-PR algorithm ($M = 2$) using different combinations of z distances as input, whereas the input low-resolution holograms were captured at fixed $z_1 = 380$ μm and $z_2 = 560$ μm.

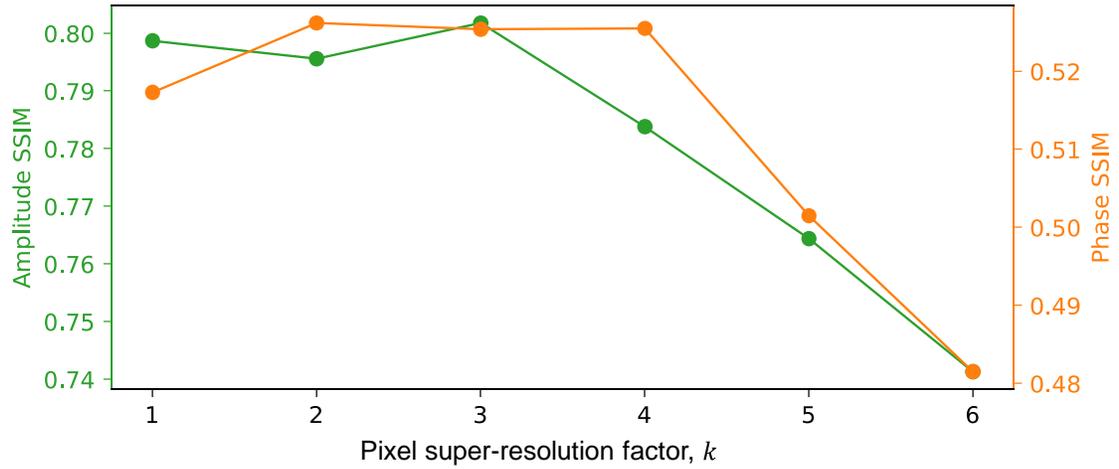

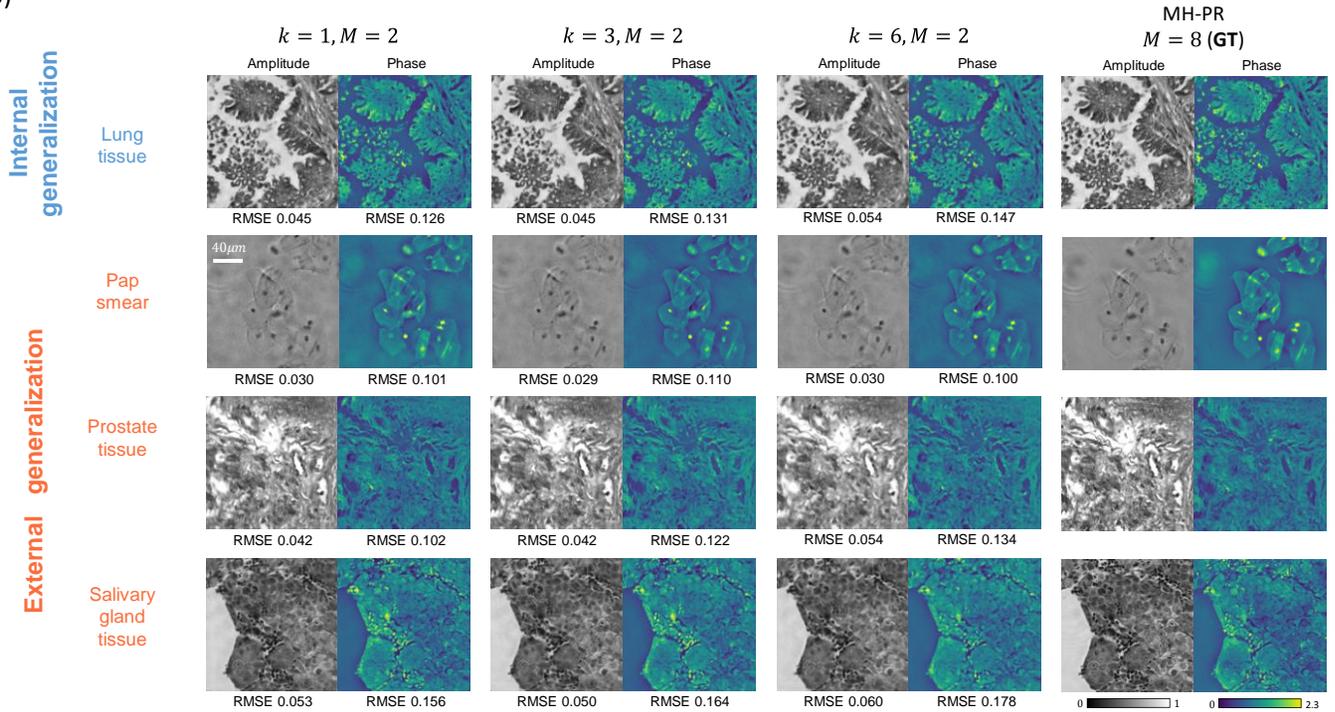

**Figure 4. Hologram reconstruction performance of eFIN as a function of the PSR factor ($k$).** An independent eFIN model ($M = 2$) was trained using human lung tissue samples for each PSR factor $k$. (a) The mean SSIM values of the amplitude and phase parts of the reconstructed sample fields for different types of samples captured at different combinations of axial distances. (b) The visualization of the eFIN reconstructed sample fields. The ground truth for each sample is obtained through the MH-PR algorithm that used $M = 8$ pixel super-resolved holograms captured at 8 different sample-to-sensor distances.

10